# The relativistic uniform model: the metric of the covariant theory of gravitation inside a body

Sergey G. Fedosin

PO box 614088, Sviazeva str. 22-79, Perm, Perm Krai, Russia

E-mail: fedosin@hotmail.com

It is shown that the sum of stress-energy tensors of the electromagnetic and gravitational fields, the acceleration field and the pressure field inside a stationary uniform spherical body within the framework of relativistic uniform model vanishes. This fact significantly simplifies solution of equation for the metric in covariant theory of gravitation (CTG). The metric tensor components are calculated inside the body, and on its surface they are combined with the components of external metric tensor. This also allows us to exactly determine one of the two unknown coefficients in the metric outside the body. Comparing the CTG metric and the Reissner-Nordström metric in general theory of relativity shows their difference, which is a consequence of difference between equations for the metric and different understanding of essence of cosmological constant.

## 1. Introduction

In modern physics, the spacetime metric of a certain physical system is completely defined by the corresponding metric tensor. The metric is of particular importance in the general theory of relativity, where the metric describes the action of gravitation. In contrast, in the covariant theory of gravitation (CTG), gravitation is an independent physical interaction. In this case, the metric of CTG is required mainly to describe the additional effects, associated with the interaction of electromagnetic waves with the gravitational field in the processes of space-time measurements by means of these waves. Accordingly, the form of the metric depends significantly on the theory of gravitation used.

Despite the success of the general theory of relativity in describing various gravitational phenomena, the theoretical foundation of this theory is still unsatisfactory. First of all, this is due to the absence of a generally recognized energy-momentum tensor of the gravitational field itself, the search for which continues to this day [1-3]. Accordingly, the energy and momentum of a system becomes ambiguous or not conserved [4-6]. Other problems include emerging singularities, the need to interpret the cosmological constant, dark matter, dark energy, etc. In this regard, the search for alternatives to the general theory of relativity remains relevant, in particular, among vector-tensor theories [7-9].

The covariant theory of gravity (CTG) refers to vector theories and has a well-defined energy-momentum tensor of the gravitational field. Outside the fixed spherical body, the metric tensor components within the framework of CTG were determined in [10]. Only the gravitational and electromagnetic fields exist outside the body, therefore only these fields exert their influence on the spacetime metric here. Using this metric, it was possible to calculate the Pioneer effect, which has no explanation in the general theory of relativity [11]. CTG formulas describing the gravitational time dilation, the gravitational redshift of the wavelength, the signal delay in the gravitational field, lead to the same results as the general theory of relativity [12].

Next, we will calculate the metric of CTG inside a spherical body. In the presence of the matter, we should take into account the pressure field, which we consider in a covariant form as a vector field. Similarly, the concept of the vector acceleration field [13-14] is used to calculate the energy and momentum of the matter, and its contribution into the equation for the metric. It is the representation of these fields in the form of vector fields that made it possible to find a covariant expression for the Navier-Stokes equation [15]. In contrast, in general relativity, the pressure field and the acceleration field are almost always considered as simple scalar fields. Consequently, we can assume that CTG more accurately represents the contribution of the fields to the energy and momentum, as well as to the metric of the system.

In order to simplify the solution of the problem, we will assume that the matter of the body moves chaotically in the volume of the spherical shape, and is kept from disruption by gravitation. The force of gravitation in such macroscopic objects, as planets and stars, is so strong that it is sufficient to form the spherical shape of these objects. This force is counteracted by the pressure force in the matter and the force from the acceleration field. One of the manifestations of the force from the acceleration field is the centrifugal force arising from that component of the particles' velocity, which is perpendicular to the radius-vector of the particles. We can also take into account the electromagnetic field and the corresponding force, which usually leads to repulsion of the charged matter in case of the excess charge of one sign. We will also assume that the physical system under consideration is a relativistic uniform system, in which the mass and charge distributions are similar to each other. This will allow us to use the expressions found earlier for the potentials and field strengths.

The need to determine the metric inside the matter arises as a consequence of the fact that the comparison of expressions for the components of the metric tensor inside and outside the matter makes it possible to unambiguously determine one of the unknown coefficients in the external metric. As a result, we obtain a more accurate expression for the CTG metric, suitable for solving more complex problems and considering small gravitational effects.

## 2. The equation for the metric

The use of the principle of least action leads to the following equation for the metric in CTG [14]:

$$R_\alpha^\beta - \frac{1}{4} R \delta_\alpha^\beta = -\frac{1}{2ck}\left(U_\alpha^\beta + W_\alpha^\beta + B_\alpha^\beta + P_\alpha^\beta\right). \tag{1}$$

Here $c$ is the speed of light; $k$ is the constant, which is part of the Lagrangian in the terms with the scalar curvature $R$ and with the cosmological constant $\Lambda$; $R_\alpha^\beta$ is the Ricci tensor with the mixed indices; $\delta_\alpha^\beta$ is the unit tensor or the Kronecker symbol; $U_\alpha^\beta$, $W_\alpha^\beta$, $B_\alpha^\beta$ and $P_\alpha^\beta$ are the stress-energy tensors of the gravitational and electromagnetic fields, the acceleration field and the pressure field, respectively.

As was shown in [16], all the quantities in (1) should be averaged over the volume of the system's typical particles, if (1) is used to find the metric inside the body. We will further assume that such averaging has already been carried out in (1). Another conclusion in [16] is that within the framework of the relativistic uniform model the scalar curvature inside a stationary body with the constant relativistically invariant mass density and charge is a certain constant quantity $\bar{R}$. In this case, in CTG the relation $\bar{R} = 2\bar{\Lambda}$ holds, where $\bar{\Lambda}$ is the averaged cosmological constant for the matter inside the body.

Acting as in [10], we will use the spherical coordinates $x^0 = ct$, $x^1 = r$, $x^2 = \theta$, $x^3 = \phi$, related to the Cartesian coordinates by the relations: $x = r\sin\theta\cos\phi$, $y = r\sin\theta\sin\phi$, $z = r\cos\theta$. For the static metric, the standard form of the metric tensor of the spherical uniform body is as follows:

$$g_{\alpha k} = \begin{pmatrix} B & 0 & 0 & 0 \\ 0 & -K & 0 & 0 \\ 0 & 0 & -E & 0 \\ 0 & 0 & 0 & -E\sin^2\theta \end{pmatrix}, \qquad (2)$$

$$g^{k\beta} = \begin{pmatrix} \dfrac{1}{B} & 0 & 0 & 0 \\ 0 & -\dfrac{1}{K} & 0 & 0 \\ 0 & 0 & -\dfrac{1}{E} & 0 \\ 0 & 0 & 0 & -\dfrac{1}{E\sin^2\theta} \end{pmatrix}, \qquad (3)$$

where $B, K, E$ are the functions of the radial coordinate $r$ only and do not depend on the angular variables, and there are four nonzero components of the metric tensor $g_{00} = B$, $g_{11} = -K$, $g_{22} = -E$, $g_{33} = -E\sin^2\theta$.

By definition, the Christoffel coefficients $\Gamma^{\beta}_{\mu\nu}$ are expressed in terms of the metric tensor and its derivatives:

$$\Gamma^{\beta}_{\mu\nu} = \frac{1}{2} g^{\beta\gamma} \left( \partial_\mu g_{\gamma\nu} + \partial_\nu g_{\gamma\mu} - \partial_\gamma g_{\mu\nu} \right). \qquad (4)$$

If we denote the derivatives with respect to the radius $r$ by primes, then the nonzero Christoffel coefficients, expressed in terms of the functions $B, K, E$ in the metric tensor (2) and (3), are equal according to (4) to the following:

$$\Gamma^0_{01} = \Gamma^0_{10} = \frac{B'}{2B}, \quad \Gamma^1_{00} = \frac{B'}{2K}, \quad \Gamma^1_{11} = \frac{K'}{2K}, \quad \Gamma^1_{22} = -\frac{E'}{2K}, \quad \Gamma^1_{33} = -\frac{E'\sin^2\theta}{2K},$$

$$\Gamma^2_{12} = \Gamma^2_{21} = \Gamma^3_{13} = \Gamma^3_{31} = \frac{E'}{2E}, \qquad \Gamma^2_{33} = -\sin\theta\cos\theta, \qquad \Gamma^3_{23} = \Gamma^3_{32} = \operatorname{ctg}\theta. \qquad (5)$$

With the help of (5) we will calculate the components of the Ricci tensor with the covariant indices using the standard formula:

$$R_{\mu\nu} = \partial_\alpha \Gamma^\alpha_{\mu\nu} - \partial_\nu \Gamma^\alpha_{\mu\alpha} + \Gamma^\alpha_{\mu\nu} \Gamma^\beta_{\alpha\beta} - \Gamma^\alpha_{\mu\beta} \Gamma^\beta_{\alpha\nu}.$$

This will give four nonzero components:

$$R_{00} = \frac{B''}{2K} - \frac{B'^2}{4BK} - \frac{B'K'}{4K^2} + \frac{B'E'}{2KE}, \qquad R_{11} = -\frac{B''}{2B} + \frac{B'^2}{4B^2} - \frac{E''}{E} + \frac{E'^2}{2E^2} + \frac{B'K'}{4BK} + \frac{K'E'}{2KE},$$

$$R_{22} = -\frac{E''}{2K} + \frac{E'K'}{4K^2} - \frac{E'B'}{4BK} + 1, \qquad R_{33} = \sin^2\theta \, R_{22}. \tag{6}$$

Equation (1) contains the components of the Ricci tensor with the mixed indices, which can be found by multiplying the components of this tensor with the covariant indices by the metric tensor using the formula: $R_\alpha^\beta = R_{\alpha\mu} g^{\mu\beta}$. With the help of (6) and (3), we find:

$$R_0^{\ 0} = \frac{B''}{2BK} - \frac{B'^2}{4B^2K} - \frac{B'K'}{4BK^2} + \frac{B'E'}{2BKE}, \qquad R_1^{\ 1} = \frac{B''}{2BK} - \frac{B'^2}{4B^2K} - \frac{B'K'}{4BK^2} + \frac{E''}{KE} - \frac{E'^2}{2KE^2} - \frac{K'E'}{2K^2E},$$

$$R_2^{\ 2} = \frac{E''}{2KE} + \frac{B'E'}{4BKE} - \frac{K'E'}{4K^2E} - \frac{1}{E}, \qquad R_3^{\ 3} = R_2^{\ 2}. \tag{7}$$

Using (6) and (3), we will calculate the scalar curvature as follows:

$$R = R_{\mu\nu} g^{\mu\nu} = \frac{B''}{BK} - \frac{B'^2}{2B^2K} - \frac{B'K'}{2BK^2} + \frac{B'E'}{BKE} + \frac{2E''}{KE} - \frac{E'^2}{2KE^2} - \frac{K'E'}{K^2E} - \frac{2}{E}. \tag{8}$$

### 3. The field tensors

The stress-energy tensors of the gravitational field [17-18], the electromagnetic field, the acceleration field and the pressure field [14], located on the right-hand side of the equation for the metric (1), can be expressed as follows

$$U_\alpha^{\ \beta} = -\frac{c^2}{4\pi G} g^{\mu\kappa}\left(-\delta_\alpha^\lambda g^{\sigma\beta} + \frac{1}{4}\delta_\alpha^\beta g^{\sigma\lambda}\right)\Phi_{\mu\lambda}\Phi_{\kappa\sigma},$$

$$W_\alpha^{\ \beta} = \varepsilon_0 c^2 g^{\mu\kappa}\left(-\delta_\alpha^\lambda g^{\sigma\beta} + \frac{1}{4}\delta_\alpha^\beta g^{\sigma\lambda}\right)F_{\mu\lambda}F_{\kappa\sigma}.$$

$$B_\alpha^{\ \beta} = \frac{c^2}{4\pi\eta} g^{\mu\kappa}\left(-\delta_\alpha^\lambda g^{\sigma\beta} + \frac{1}{4}\delta_\alpha^\beta g^{\sigma\lambda}\right)u_{\mu\lambda}u_{\kappa\sigma},$$

$$P_\alpha^{\ \beta} = \frac{c^2}{4\pi\sigma} g^{\mu\kappa}\left(-\delta_\alpha^\lambda g^{\sigma\beta} + \frac{1}{4}\delta_\alpha^\beta g^{\sigma\lambda}\right)f_{\mu\lambda}f_{\kappa\sigma}. \tag{9}$$

In (9) $\Phi_{\mu\lambda}$, $F_{\mu\lambda}$, $u_{\mu\lambda}$ and $f_{\mu\lambda}$ are the tensors of the gravitational and electromagnetic fields, the acceleration field and the pressure field, respectively; $G$ is the gravitational constant; $\varepsilon_0$ is the electric constant (vacuum permittivity); $\eta$ is the acceleration field constant; $\sigma$ is the pressure field constant.

The stress-energy tensors in (9) were derived from the principle of least action under the assumption that all the physical fields in the system under consideration were described as vector fields that had their own four-potentials [13]. Due to the fact that the field tensors have the same form, it was possible to combine all the fields into a single general field [19-20].

Let us express the four-potentials of the fields in terms of the corresponding scalar and vector potentials of these fields: $D_\lambda = \left(\frac{\psi}{c}, -\mathbf{D}\right)$ for the gravitational field, $A_\lambda = \left(\frac{\varphi}{c}, -\mathbf{A}\right)$ for the electromagnetic field, $U_\lambda = \left(\frac{\vartheta}{c}, -\mathbf{U}\right)$ for the acceleration field, and $\pi_\lambda = \left(\frac{\wp}{c}, -\mathbf{\Pi}\right)$ for the pressure field. The gravitational tensor is defined as the four-curl of the four-potential [17]. Similarly, the electromagnetic tensor, the acceleration tensor and the pressure field tensor [14] are calculated:

$$\Phi_{\mu\lambda} = \nabla_\mu D_\lambda - \nabla_\lambda D_\mu = \partial_\mu D_\lambda - \partial_\lambda D_\mu, \qquad F_{\mu\lambda} = \nabla_\mu A_\lambda - \nabla_\lambda A_\mu = \partial_\mu A_\lambda - \partial_\lambda A_\mu.$$

$$u_{\mu\lambda} = \nabla_\mu U_\lambda - \nabla_\lambda U_\mu = \partial_\mu U_\lambda - \partial_\lambda U_\mu, \qquad f_{\mu\lambda} = \nabla_\mu \pi_\lambda - \nabla_\lambda \pi_\mu = \partial_\mu \pi_\lambda - \partial_\lambda \pi_\mu.$$
$$\tag{10}$$

In the system under consideration, the vector potentials **D**, **A**, **U** and **Π** of all the fields are close to zero because of the random motion of the matter's particles. This is due to the fact that the vector potentials of individual particles are directed along the particles' velocities, and therefore they change each time as a result of interactions. The global vector potential of each field inside the body is calculated as the vector sum of the corresponding vector potentials of the particles. At each time point, most of the particles in the system have oppositely directed velocities and vector potentials, so that on the average the vector sum of these potentials tends to zero. The more particles are present in the system, the more exactly the equality to zero holds for the global vector potentials of the fields. We will not also take into account the proper vector potentials of individual particles. As was shown in [21], the energy of the particles' motion arises due to all these potentials, which is approximately equal to their kinetic energy. Thus the inaccuracy, arising from equating the vector potentials **D**, **A**, **U** and **Π** to zero, does not exceed the inaccuracy in the case when only the rest energy is taken into account in the system's energy and the kinetic energy of the particles is neglected.

As for the scalar field potentials $\psi$, $\varphi$, $\vartheta$ and $\wp$, in the static case for a stationary spherical body they must depend only on the current radius $r$ and must not depend on either time or angular variables.

Assuming that $\mathbf{D} \approx 0$ and neglecting the contribution of the vector potential **D**, in the spherical coordinates $x^0 = ct$, $x^1 = r$, $x^2 = \theta$, $x^3 = \phi$ from (10) and (3) we find the nonzero components of the gravitational tensor:

$$\Phi_{01} = -\Phi_{10} = -\frac{1}{c}\frac{\partial \psi}{\partial r} - \frac{1}{c}\frac{\partial D_r}{\partial t} \approx -\frac{1}{c}\frac{\partial \psi}{\partial r}. \tag{11}$$

In (11) the quantity $D_r$ in the spherical coordinates is the projection of the vector potential on the radial component of the four-dimensional coordinate system. In this case, the quantity $\Gamma_r = c\Phi_{01} = -\frac{\partial \psi}{\partial r} - \frac{\partial D_r}{\partial t}$ is the projection of the gravitational field strength on the radial component of the coordinate system.

The nonzero components of the electromagnetic tensor, the acceleration tensor, and the pressure field tensor are obtained similarly to (11):

$$F_{01} = -F_{10} = -\frac{1}{c}\frac{\partial \varphi}{\partial r} - \frac{1}{c}\frac{\partial A_r}{\partial t} \approx -\frac{1}{c}\frac{\partial \varphi}{\partial r}. \qquad u_{01} = -u_{10} = -\frac{1}{c}\frac{\partial \vartheta}{\partial r} - \frac{1}{c}\frac{\partial U_r}{\partial t} \approx -\frac{1}{c}\frac{\partial \vartheta}{\partial r}.$$

$$f_{01} = -f_{10} = -\frac{1}{c}\frac{\partial \wp}{\partial r} - \frac{1}{c}\frac{\partial \Pi_r}{\partial t} \approx -\frac{1}{c}\frac{\partial \wp}{\partial r}. \qquad (12)$$

In Minkowski spacetime the special theory of relativity is valid, so that the potentials and the field strengths can be calculated exactly. For the case of the relativistic uniform model, the field strengths, which are part of the field tensors' components inside a spherical body, in the static case have the following form [22]:

$$c\Phi_{01} = -\frac{Gc^2\gamma_c}{\eta r^2}\left[\frac{c}{\sqrt{4\pi\eta\rho_0}}\sin\left(\frac{r}{c}\sqrt{4\pi\eta\rho_0}\right) - r\cos\left(\frac{r}{c}\sqrt{4\pi\eta\rho_0}\right)\right] \approx -\frac{4\pi G\rho_0\gamma_c r}{3},$$

$$cF_{01} = \frac{\rho_{0q}c^2\gamma_c}{4\pi\varepsilon_0\rho_0\eta r^2}\left[\frac{c}{\sqrt{4\pi\eta\rho_0}}\sin\left(\frac{r}{c}\sqrt{4\pi\eta\rho_0}\right) - r\cos\left(\frac{r}{c}\sqrt{4\pi\eta\rho_0}\right)\right] \approx \frac{\rho_{0q}\gamma_c r}{3\varepsilon_0},$$

$$cu_{01} = \frac{c^2\gamma_c}{r^2}\left[\frac{c}{\sqrt{4\pi\eta\rho_0}}\sin\left(\frac{r}{c}\sqrt{4\pi\eta\rho_0}\right) - r\cos\left(\frac{r}{c}\sqrt{4\pi\eta\rho_0}\right)\right] \approx \frac{4\pi\eta\rho_0\gamma_c r}{3},$$

$$cf_{01} = \frac{\sigma c^2\gamma_c}{\eta r^2}\left[\frac{c}{\sqrt{4\pi\eta\rho_0}}\sin\left(\frac{r}{c}\sqrt{4\pi\eta\rho_0}\right) - r\cos\left(\frac{r}{c}\sqrt{4\pi\eta\rho_0}\right)\right] \approx \frac{4\pi\sigma\rho_0\gamma_c r}{3}.$$

(13)

In (13) $\gamma_c$ is the Lorentz factor of the typical particles that are moving in the center of the body; $\rho_0$ and $\rho_{0q}$ denote the invariant mass and charge densities of the typical particles, respectively. These mass and charge densities are obtained in the reference frames, which are comoving with the particles. It follows from (10-13) that the field tensors inside the body are proportional to each other:

$$-\frac{\Phi_{\mu\lambda}}{G} = \frac{4\pi\varepsilon_0\rho_0 F_{\mu\lambda}}{\rho_{0q}} = \frac{u_{\mu\lambda}}{\eta} = \frac{f_{\mu\lambda}}{\sigma}. \qquad (14)$$

Let us sum up all the stress-energy tensors in (9) and use (14):

$$U_\alpha^{\ \beta} + W_\alpha^{\ \beta} + B_\alpha^{\ \beta} + P_\alpha^{\ \beta} = \frac{c^2}{4\pi\eta^2}\left(-G + \frac{\rho_{0q}^2}{4\pi\varepsilon_0 \rho_0^2} + \eta + \sigma\right) g^{\mu\kappa}\left(-\delta_\alpha^\lambda g^{\sigma\beta} + \frac{1}{4}\delta_\alpha^\beta g^{\sigma\lambda}\right) u_{\mu\lambda} u_{\kappa\sigma}.$$

(15)

As was found in [23] from the equation of the particles' motion and in [24] from the generalized Poynting theorem, the following condition holds for the sum of the field coefficients inside the body:

$$-G + \frac{\rho_{0q}^2}{4\pi\varepsilon_0 \rho_0^2} + \eta + \sigma = 0.$$

(16)

Substituting condition (16) into (15) we find out that the sum of the stress-energy tensors inside the body, which is in equilibrium, becomes equal to zero:

$$U_\alpha^{\ \beta} + W_\alpha^{\ \beta} + B_\alpha^{\ \beta} + P_\alpha^{\ \beta} = 0.$$

(17)

Will the result of (17) change if we consider the situation in the curved spacetime? In the physical system under consideration in the form of a spherical body, the spacetime metric is static and depends only on the radial coordinate. Since the vector field potentials are assumed to be zero, the tensor of each field contains only two nonzero components, which are equal in the absolute value. Taking into account the metric of the curved spacetime leads to the fact that the tensors' components of each field in (13) must be multiplied by the same function $Z$ that depends on the metric tensor components. Just as the metric tensor components, this function will depend only on the radial coordinate. In this case, in the flat Minkowski spacetime this function must be equal to unity, $Z = 1$, so that (13) is satisfied, which does not contain $Z$.

Indeed, the equations for calculating the tensors of all the vector fields coincide with each other in their form, according to [13], [18], [25], so that in the relativistic uniform model at constant mass density $\rho_0$ and charge density $\rho_{0q}$ the field tensors can differ from each other only by the constant coefficients. Therefore, if we multiply the tensor of each field, found in the Minkowski spacetime, by the same function $Z$, in order to find this tensor in the curved spacetime, relation (14) would not change, and an additional factor $Z^2$ would appear on the right-hand side of (15). Since condition (16) always holds true, then in the system under consideration the sum of the stress-energy tensors in (15) and (17) will also be zero in the curved spacetime.

### 4. Calculation of the metric inside the body

Equation for the metric (1) in view of (17) is significantly simplified:

$$R_\alpha^{\ \beta} - \frac{1}{4} R \delta_\alpha^{\ \beta} = 0.$$

Substituting here (7) and (8), we obtain three equations:

$$\frac{B''}{2BK} - \frac{B'^2}{4B^2K} - \frac{B'K'}{4BK^2} + \frac{B'E'}{2BKE} - \frac{E''}{KE} + \frac{E'^2}{4KE^2} + \frac{K'E'}{2K^2E} + \frac{1}{E} = 0. \quad (18)$$

$$\frac{B''}{2BK} - \frac{B'^2}{4B^2K} - \frac{B'K'}{4BK^2} - \frac{B'E'}{2BKE} + \frac{E''}{KE} - \frac{3E'^2}{4KE^2} - \frac{K'E'}{2K^2E} + \frac{1}{E} = 0. \quad (19)$$

$$\frac{B''}{2BK} - \frac{B'^2}{4B^2K} - \frac{B'K'}{4BK^2} - \frac{E'^2}{4KE^2} + \frac{1}{E} = 0. \quad (20)$$

Substituting (20) in (18) and in (19) gives the same equation:

$$\frac{E''}{KE} - \frac{E'^2}{2KE^2} - \frac{B'E'}{2BKE} - \frac{K'E'}{2K^2E} = 0, \quad \text{or} \quad \frac{E''}{E'} - \frac{E'}{2E} - \frac{B'}{2B} - \frac{K'}{2K} = 0. \quad (21)$$

If we subtract equation (18) from (19), we will obtain again (21). Equation (21) can be easily integrated, because each term represents the derivative of the natural logarithm of the corresponding function:

$$E' = C_1 \sqrt{BKE}, \quad (22)$$

where $C_1$ is a certain constant.

We will now use the condition obtained in [16], according to which the scalar curvature inside the body must be a constant value $\bar{R} = C_2$. With the help of (8) we find:

$$\frac{B''}{BK} - \frac{B'^2}{2B^2K} - \frac{B'K'}{2BK^2} + \frac{B'E'}{BKE} + \frac{2E''}{KE} - \frac{E'^2}{2KE^2} - \frac{K'E'}{K^2E} - \frac{2}{E} = C_2. \qquad (23)$$

The sum of (23) and (18) gives the following:

$$\frac{B''}{BK} - \frac{B'^2}{2B^2K} - \frac{B'K'}{2BK^2} + \frac{B'E'}{BKE} = \frac{C_2}{2}.$$

Comparing this expression with (20), we obtain:

$$\frac{E'^2}{4KE^2} - \frac{1}{E} + \frac{B'E'}{2BKE} = \frac{C_2}{4}.$$

We will substitute here the value of $K = \dfrac{E'^2}{C_1^2 BE}$ according to (22):

$$\frac{E'}{E} = \frac{2C_1^2 B'}{4 + C_2 E - C_1^2 B}. \qquad (24)$$

Next we will need $K$ from (22) and the relation $\dfrac{K'}{K}$ from (21):

$$K = \frac{E'^2}{C_1^2 BE}, \qquad \frac{K'}{K} = \frac{2E''}{E'} - \frac{B'}{B} - \frac{E'}{E}. \qquad (25)$$

Let us substitute (25) into (20):

$$B'' - \frac{B'E''}{E'} + \frac{B'E'}{2E} - \frac{BE'^2}{2E^2} + \frac{2E'^2}{C_1^2 E^2} = 0. \qquad (26)$$

Equations (24) and (26) together form a system of two differential equations for the two functions $B$ and $E$. Direct substitution shows us that the system of these equations has the following solution: $E = r^2$, $B = \dfrac{C_3}{r} + \dfrac{4}{C_1^2} + \dfrac{C_2 r^2}{3 C_1^2}$. Indeed, in the weak gravitational field, when the curved spacetime turns into the Minkowski spacetime, in the spherical coordinates it should be $E = r^2$, $B = K \approx 1$. In order to ensure that the function $B$ is not infinitely large at the center at $r = 0$, the constant $C_3$ must be equal to zero. From the condition $B \approx 1$ it follows that $C_1 = 2$, and from (22) we obtain the equality $BK = 1$. In addition, the constant $C_2$ must be sufficiently small. As a result, for the metric tensor components we can write the following:

$$B = g_{00} = 1 + \frac{C_2 r^2}{12}, \qquad E = -g_{22} = r^2,$$

$$K = -g_{11} = \frac{1}{B} = \frac{1}{1 + \dfrac{C_2 r^2}{12}}, \qquad g_{33} = -r^2 \sin^2 \theta. \qquad (27)$$

The constant $C_2$ in (27) represents the value of the scalar curvature, averaged over the volume of a typical particle, which is constant inside the body, so that $\bar{R} = C_2$.

In [16] we found the relation for the value of the cosmological constant $\bar{\Lambda}$ averaged over the volume of a typical particle:

$$-ck\bar{\Lambda} = \frac{G\rho_0}{a}\left[\frac{c^3 \gamma_c}{\eta\sqrt{4\pi\eta\rho_0}}\sin\left(\frac{a}{c}\sqrt{4\pi\eta\rho_0}\right) - m_g\right] -$$

$$-\frac{\rho_{0q}}{4\pi\varepsilon_0 a}\left[\frac{\rho_{0q} c^3 \gamma_c}{\eta\rho_0\sqrt{4\pi\eta\rho_0}}\sin\left(\frac{a}{c}\sqrt{4\pi\eta\rho_0}\right) - q_b\right] + \rho_0\wp_c - \frac{\sigma\rho_0 c^2 \gamma_c}{\eta}.$$

Expanding the sine by the rule $\sin x \approx x - \dfrac{x^3}{6}$, in view of (16), we find:

$$-ck\bar{\Lambda} \approx -\frac{Gm_g\rho_0}{a} - \frac{Gm\rho_0\gamma_c}{2a} + \rho_0 c^2 \gamma_c + \frac{q_b\rho_{0q}}{4\pi\varepsilon_0 a} + \frac{q\rho_{0q}\gamma_c}{8\pi\varepsilon_0 a} + \rho_0\wp_c.$$

$$C_2 = \bar{R} = 2\bar{\Lambda} \approx -\frac{2}{ck}\left(\rho_0\psi_a - \frac{Gm\rho_0\gamma_c}{2a} + \rho_0 c^2\gamma_c + \rho_{0q}\varphi_a + \frac{q\rho_{0q}\gamma_c}{8\pi\varepsilon_0 a} + \rho_0\wp_c\right). \tag{28}$$

In (28) $k = -\dfrac{c^3}{16\pi G\beta}$, where $\beta$ is a certain constant of the order of unity; $\psi_a = -\dfrac{Gm_g}{a}$ is the scalar potential of the gravitational field on the surface of the body at $r = a$; $\varphi_a = \dfrac{q_b}{4\pi\varepsilon_0 a}$ is the scalar potential of the electric field; $a$ is the radius of the body; $m_g$ is the gravitational mass of the body; $q_b$ is the total charge of the body; $\gamma_c$ is the Lorentz factor of the particles at the center of the body; $\wp_c$ is the potential of the pressure field at the center of the sphere; the mass $m = \dfrac{4\pi a^3\rho_0}{3}$ and the charge $q = \dfrac{4\pi a^3\rho_{0q}}{3}$ are auxiliary quantities.

In the brackets on the right-hand side of (28) there is the sum of the volumetric energy densities of the particles in the scalar field potentials – the first and second terms are from the gravitational field, the third term is from the acceleration field, the fourth and fifth terms are from the electric field, and the sixth term is from the pressure field. The third term is the greatest, it is proportional to the rest energy density of the body. If we take into account only this term, then in the first approximation the constant $C_2$ will be equal to:

$$C_2 \approx \frac{32\pi G\rho_0\gamma_c\beta}{c^2}.$$

**5. Comparison of the metric tensor components inside and outside the body**

At $r = a$ the current radius reaches the surface of the spherical body, and here the internal metric becomes equal to the external metric. It means that at $r = a$ we can equate the components of the corresponding metric tensors. According to [10], the metric tensor components outside the body in the covariant theory of gravitation are equal to:

$$g_{00} = 1 + \frac{A_3}{r} + \frac{1}{16\pi ckr^2}\left(Gm_g^2 - \frac{q_b^2}{4\pi\varepsilon_0}\right), \qquad g_{22} = -r^2,$$

$$g_{11} = -\frac{1}{1 + \dfrac{A_3}{r} + \dfrac{1}{16\pi ckr^2}\left(Gm_g^2 - \dfrac{q_b^2}{4\pi\varepsilon_0}\right)}, \qquad g_{33} = -r^2 \sin^2\theta. \qquad (29)$$

Comparison of (29) and (27) shows that the components $g_{22}$ and $g_{33}$ coincide both inside and outside the body.

Equating $g_{00}$ in (27) and (29) on condition that $r = a$, taking into account (28) and the equality $k = -\dfrac{c^3}{16\pi G\beta}$, we find the constant $A_3$:

$$A_3 \approx \frac{2G\beta}{c^4}\left[mc^2\gamma_c + \left(m - \frac{1}{2}m_g\right)\psi_a - \frac{Gm^2\gamma_c}{2a} + \left(q - \frac{1}{2}q_b\right)\varphi_a + \frac{q^2\gamma_c}{8\pi\varepsilon_0 a} + m\wp_c\right]. \qquad (30)$$

According to [22], the gravitational mass $m_g$ of the body and the total electric charge $q_b$ are determined as follows:

$$m_g = \int \rho_0 \gamma' dV = \frac{c^2\gamma_c}{\eta}\left[\frac{c}{\sqrt{4\pi\eta\rho_0}}\sin\left(\frac{a}{c}\sqrt{4\pi\eta\rho_0}\right) - a\cos\left(\frac{a}{c}\sqrt{4\pi\eta\rho_0}\right)\right] \approx m\gamma_c\left(1 - \frac{3\eta m}{10ac^2}\right).$$

(31)

$$q_b = \int \rho_{0q}\gamma' dV =$$
$$= \frac{\rho_{0q}c^2\gamma_c}{\eta\rho_0}\left[\frac{c}{\sqrt{4\pi\eta\rho_0}}\sin\left(\frac{a}{c}\sqrt{4\pi\eta\rho_0}\right) - a\cos\left(\frac{a}{c}\sqrt{4\pi\eta\rho_0}\right)\right] \approx q\gamma_c\left(1 - \frac{3\eta m}{10ac^2}\right).$$

Since $\gamma_c > 0$, it turns out that $m_g > m$ and $q_b > q$.

We will substitute (30) into (29) into the expression for $g_{00}$ and take into account the equality $k = -\dfrac{c^3}{16\pi G\beta}$:

$$g_{00} = -\frac{1}{g_{11}} \approx 1 + \frac{2Gm\gamma_c \beta}{c^2 r} +$$
$$+ \frac{2G\beta}{c^4 r}\left[ m\psi_a + \frac{1}{2}m_g(\psi - \psi_a) - \frac{Gm^2 \gamma_c}{2a} + q\varphi_a + \frac{1}{2}q_b(\varphi - \varphi_a) + \frac{q^2 \gamma_c}{8\pi\varepsilon_0 a} + m\wp_c \right].$$
(32)

In this expression $\psi = -\frac{Gm_g}{r}$ and $\varphi = \frac{q_b}{4\pi\varepsilon_0 r}$ denote the scalar potentials of the gravitational and electric fields outside the body, respectively. We can also determine the quantities $\gamma_c$ and $\wp_c$ more exactly. In [21] we found the expression for the square of the particles' velocities $v_c^2$ at the center of the spherical body, with the help of which we can estimate the value of the Lorentz factor in (32):

$$\gamma_c = \frac{1}{\sqrt{1-v_c^2/c^2}} \approx 1 + \frac{v_c^2}{2c^2} + \frac{3v_c^4}{8c^4} \approx 1 + \frac{3\eta m}{10ac^2}\left(1 + \frac{9}{2\sqrt{14}}\right) + \frac{27\eta^2 m^2}{200a^2 c^4}\left(1 + \frac{9}{2\sqrt{14}}\right)^2.$$

According to [26], the scalar potential of the pressure field at the center of the body is approximately equal to:

$$\wp_c \approx \frac{3\sigma m}{10a}\left(1 + \frac{9}{2\sqrt{14}}\right),$$

while the acceleration field constant $\eta$ and the pressure field constant $\sigma$ are given by the formulas:

$$\eta = \frac{3}{5}\left(G - \frac{\rho_{0q}^2}{4\pi\varepsilon_0 \rho_0^2}\right), \qquad \sigma = \frac{2}{5}\left(G - \frac{\rho_{0q}^2}{4\pi\varepsilon_0 \rho_0^2}\right).$$

In (32) we see the complex structure of the metric tensor components, in which additional terms appear as compared to the Minkowski spacetime metric, where in the spherical coordinates $g_{00} = -\frac{1}{g_{11}} = 1$. The main addition in (32) is the term $\frac{2Gm\gamma_c \beta}{c^2 r}$, and if we take into account (31), then this addition will become approximately equal to $-\frac{2\psi\beta}{c^2}$.

The second important addition includes square brackets in (32), which by the order of magnitude determines the energy of the gravitational and electric fields, as well as the pressure field. In these brackets, we can also use the approximate relation of the masses in expression (31). For the metric tensor components outside the body all this leads to the following:

$$g_{00} = -\frac{1}{g_{11}} \approx 1 - \frac{2\psi\beta}{c^2} + \frac{2G\beta}{c^4 r}\left(m\psi_a + \frac{1}{2}m_g\psi + \frac{3\eta m^2 \gamma_c}{10a} + q\varphi_a + \frac{1}{2}q_b\varphi + m\wp_c\right). \quad (33)$$

On the right-hand side of (33) in the round brackets there are quantities with the dimension of energy. For large cosmic bodies, the main quantity here is the negative energy associated with gravitation. In this case we can see that the third term, containing $c^4$ in the denominator, is distinguished by a sign from the second term, containing $c^2$ in the denominator.

### 6. Comparison with the metric of the general theory of relativity

In order to compare with the metric tensor components (29) and (33), we will consider the Reissner-Nordström metric in the spherical coordinates, which describes the static gravitational field around a charged spherical body in the general theory of relativity. We will use our notation for the field potentials:

$$g_{00} = 1 + \frac{2\psi}{c^2} + \frac{Gq_b\varphi}{c^4 r}, \qquad g_{11} = -\frac{1}{1 + \frac{2\psi}{c^2} + \frac{Gq_b\varphi}{c^4 r}},$$

$$g_{22} = -r^2, \qquad g_{33} = -r^2 \sin^2\theta. \quad (34)$$

As we can see, the second and third terms in the component $g_{00}$ in the Reissner-Nordström metric (34) differ significantly from the corresponding terms in the component $g_{00}$ in the CTG metric (33) outside the body. For example, we can see that the metric in (34) does not in any way reflect the energy of the pressure field inside the body, whereas in (33) the energy $m\wp_c$ is associated with the pressure field and makes its contribution to the metric. Taking into account (28), the energy $m\wp_c$ also defines the metric (27) inside the body.

This difference in the form of the metric in both theories is due to the difference in the equations for determining the metric. While equation (1) is used in CTG, in the general theory of relativity the equation for the metric with the cosmological constant $\Lambda$, in the matter with the stress-energy tensor $T_\alpha^{\ \beta}$ has the following form:

$$R_\alpha^{\ \beta} - \frac{1}{2} R \delta_\alpha^{\ \beta} + \Lambda \delta_\alpha^{\ \beta} = \frac{8\pi G}{c^4} T_\alpha^{\ \beta}. \tag{35}$$

According to the approach of the general theory of relativity, the action of gravitation must be described by the metric tensor, and therefore $T_\alpha^{\ \beta}$ does not include the stress-energy tensor of the gravitational field. Outside the charged body there is no matter and no pressure field, on the right-hand side of (35) only the electromagnetic field is left, so that we have $T_\alpha^{\ \beta} = W_\alpha^{\ \beta}$. As a rule, in (35) the term with the cosmological constant $\Lambda$ is neglected due to its smallness, and then the solution for the metric (34) is obtained.

Since in CTG the cosmological constant is taken into account fully, it turns out that the solution of (27) in view of (28) for the CTG metric inside the body and the solution of (32) outside the body are more precise and informative than the solution of (34) in the Reissner-Nordström metric. Besides the cosmological constant $\bar{\Lambda}$ in CTG is not equal to zero and is proportional to the potentials of all the fields acting inside the body. If in (28) only the main term with rest energy density is taken into account, then with the relation $k = -\dfrac{c^3}{16\pi G \beta}$ we can estimate the value $\bar{\Lambda}$:

$$\bar{\Lambda} \approx -\frac{\rho_0 c \gamma_c}{k} \approx \frac{16 \pi G \rho_0 \beta \gamma_c}{c^2}. \tag{36}$$

If we substitute here the average mass density of the cosmic space matter of the observable Universe, we will obtain the value $\bar{\Lambda} \approx 10^{-52}$ m$^{-2}$. The smallness of the cosmological constant $\bar{\Lambda}$ inside cosmic bodies is associated with the large factor $k = -\dfrac{c^3}{16\pi G \beta}$ in (36). To this end we recall that the issue of the cosmological constant in the general theory of relativity has not yet been resolved unambiguously [27], especially with respect to correlation with vacuum energy. Here it is implied that a very large vacuum energy for some reason makes little contribution to the metric and to the small cosmological constant.

In CTG, the greater is the mass density in (36), the larger is $\bar{\Lambda}$ inside the body. However if we distribute the matter of all cosmic bodies over the space, then the mass density will be very low, which leads to insignificantly small value $\bar{\Lambda} \approx 10^{-52}$ m$^{-2}$. We should also pay attention to the fact that in CTG the cosmological constant outside the body is assumed to be zero due to its gauging [16]. Inside the bodies, as well as inside the observable Universe as some global body, $\bar{\Lambda}$ has a certain value. In the approximation of the relativistic uniform body model, $\bar{\Lambda}$ is determined in (28).

In contrast, in the general theory of relativity, in (35), the nonzero value of the cosmological constant outside the body is admitted. The latter follows from the possibility of influence of the zero vacuum's energy on the metric through the cosmological constant.

## 7. Conclusions

In Section 3 we showed that the sum of the stress-energy tensors of all the four fields inside the body is zero. With this in mind, the metric tensor components were calculated in (27) as functions of the current radius. As a result, on the surface of the body at $r = a$ it became possible to compare the metric inside and outside the body and to determine the unknown coefficient $A_3$ in the external metric (29).

The metric tensor components $g_{00}$ and $g_{11}$ outside the fixed spherical body in the covariant theory of gravitation (CTG), that were presented in (29), were specified by us in (32) and (33). It turns out that these components are the functions of the scalar potentials of all the fields, so that, for example, the pressure field inside the body also influences the metric outside the body. However, the main contribution to the metric is made by the scalar potential of the gravitational field $\psi = -\dfrac{Gm_g}{r}$. Apparently this is due to the fact that the expression for the scalar potential $\psi$ includes the gravitational mass $m_g$ that characterizes the source of the field and the gravitation force. At the same time the relativistic energy is proportional to the inertial mass $M$, while for an external observer the mass $M$ is the rest mass and characterizes the system with respect to the forces acting on it. Both of these masses differ from each other by the mass-energy of the particles' binding by means of the fields [26]. As for the electromagnetic field, its contribution is secondary. The body's charge is only indirectly included in the rest mass of the body and is not directly included in the gravitational mass. The electric field potentials vanish in neutral bodies in (33). Thus the gravitational field is the main factor that distinguishes the curved spacetime metric from the Minkowski flat spacetime metric.

Our calculations allowed us to calculate the metric CTG inside the body and to refine the metric outside the body, but in the metric tensor components there was one more unknown adjustable coefficient $\beta$. Its appearance can be due to the assumption that the coefficient $k = -\dfrac{c^3}{16\pi G\beta}$ has an exact value, so that the coefficient $\beta$ is intended to ensure the correct value of the metric. The value of the coefficient $\beta$ can be determined in the gravitational experiments, in which the spacetime metric should be taken into account.

## 8. Conflict of interest

On behalf of all authors, the corresponding author states that there is no conflict of interest.